\documentclass[12pt]{article}
\usepackage{authblk}
\usepackage{graphicx}
\usepackage{amsmath}
\usepackage{amssymb}
\usepackage{booktabs}
\usepackage{siunitx}
\usepackage{hyperref}
\usepackage{natbib}
\usepackage{geometry}
\usepackage{setspace}
\usepackage{titlesec}
\usepackage{caption}
\usepackage{listings}
\usepackage{xcolor}
\usepackage{float}
\usepackage{listings}
\usepackage{xcolor}
\title{Signatures of Galactic Expansion in Gaia DR3:\\
Implications for the JWST Early-Galaxy Puzzle}

\author[1]{G.~S.~Karapetian}
\author[1]{A.~P.~Mahtessian*}
\author[2]
{L.~E.~Byzalov}
\author[3]{M.~A.~Hovhannisyan}
\author[3]{L.~A.~Mahtessian}

\affil[1]{Byurakan Astrophysical Observatory after V. Ambartsumian, NAS of the Republic of Armenia, Byurakan, Aragatzotn Province, Republic of Armenia, 0213. *amahtes@gmail.com}
\affil[2]{University of Waterloo, Ontario, Canada, N2L 3G1}
\affil[3]{Institute of Applied Problems of Physics, NAS of RA, Yerevan}
\date{}
\begin{document}
\maketitle
\begin{abstract}
Recent observations with the James Webb Space Telescope (JWST) of massive galaxies at ages below 1 Gyr pose a challenge to standard models of galaxy formation, which predict significantly longer assembly timescales. One possible explanation is that active galactic nuclei (AGN) drive large-scale outflows that accelerate galaxy growth. To test this scenario in the local Universe, we analyzed Gaia DR3 data for stars within 5~kpc of the Galactic center, computing galactocentric radial velocities ($v\_{\mathrm{radial} \mathrm{\_gc}}$) in 27 spatial sectors covering the entire Galaxy, with radial binning of 0.25~kpc. Coordinate transformations and velocity calculations were performed using the Astropy library. We find that 21 of 27 sectors exhibit statistically significant outward motions of $3–50 ~km~s^{-1}$, while one quadrant shows negative velocities, likely related to the configuration of an activity zone and/or the Galactic bar. Both disk and halo populations also display small but significant mean expansion of 3--9~km~s$^{-1}$ ($p<0.01$). These results are consistent with our previous studies, where globular clusters showed outward velocities of 17--31~km~s$^{-1}$ up to 12~kpc, and axisymmetric analyses of Gaia DR3 stars indicated expansion of $\sim$19~km~s$^{-1}$ to 5~kpc. Taken together, the evidence suggests that the Milky Way exhibits measurable central expansion, potentially reflecting AGN-driven feedback. This interpretation departs from standard theory and should be regarded as preliminary, requiring further study. However, if confirmed, such expansion could provide a natural explanation for the rapid appearance of massive galaxies observed by JWST.
\end{abstract}
\section{Introduction}
In standard astrophysics, galaxies are generally modeled as systems in gravitational equilibrium. According to the Jeans equations \citep{BinneyTremaine2008}, their disk, bulge, and halo can be described as steady-state structures without large-scale radial expansion or contraction. Milky Way models \citep{WidrowDubinski2005,McMillan2016} and cosmological simulations \citep{Dave2011} likewise assume equilibrium, making significant radial expansion on sub--10~kpc scales highly unlikely.

The Gaia DR3 release provides an unprecedented opportunity to test this assumption with large, high-precision stellar datasets. (Martínez-Lombilla et al. 2018) noted that inside-out disc growth is limited to 0.6-1~km~s$^{-1}$, confined to the disc edge rather than the central region. In contrast, our earlier studies found evidence for outward motions in globular clusters (17--31~km~s$^{-1}$ out to 12~kpc; \citep{Karapetian2025a}) and in stars along the south--north axis ($\sim$19~km~s$^{-1}$ out to 5~kpc; \citep{Karapetian2025b}), suggesting that the Milky Way itself may undergo large-scale expansion.
Detecting such expansion would imply that the Galactic nucleus is dynamically active and acts as a source of matter and energy, consistent with AGN-driven outflows. This possibility gains particular importance in light of JWST discoveries of massive galaxies within a few hundred Myr of the Big Bang \citep{Ferrara2022,Xiao2023,Nanayakkara2022,Carniani2024}. If galactic nuclei can drive expansion, they may provide a natural explanation for the unexpectedly rapid assembly of galaxies in the early universe.

\section{Method}
To test the hypothesis of possible Galactic expansion, we used stellar data from Gaia DR3 \citep{GaiaDR3}. To probe whether expansion is a global property, we defined 27 galactocentric sectors  (Table 1, Appendix 1) covering all quadrants of the disk, northern and southern halo regions, and the Galaxy as a whole. The disk--halo division was made using $|z|<0.5$~kpc \citep{Juric2008,GilmoreReid1983,Bovy2012,Robin2003,Yoachim2006}. The analysis was limited to galactocentric distances $r_{\mathrm{gc}}\le 5$~kpc.
Because Gaia DR3 does not provide complete six-dimensional phase-space information (positions + velocities) with uniform accuracy and coverage across all quadrants, the number of stars per sector ranges from 21 to 31,777. In our initial astroquery.gaia requests, the maximum sample size was set to 50,000 stars. The chosen set of sectors thus spans all major Galactic directions and regions, providing a representative test of the presence or absence of expansion.

Data extraction was carried out using ADQL \citep{Osuna2008} via the \texttt{astroquery.gaia} library \citep{Ginsburg2019} in Python \citep{VanRossum2009}. To ensure accuracy, the following quality filters were applied:
\begin{verbatim}
parallax > 0
AND parallax_over_error > 5
AND ABS(radial_velocity) > 0
AND ABS(radial_velocity / radial_velocity_error) > 5
AND ABS(pmra / pmra_error) > 5
AND ABS(pmdec / pmdec_error) > 5
\end{verbatim}
The complete Python code implementing these queries and transformations is provided in \textbf{Appendix 2.}
Stellar coordinates and velocities were transformed to the Galactocentric frame using \texttt{Astropy} and its \texttt{Galactocentric} model \citep{Astropy2018}. For each star we computed the galactocentric radius $r{\mathrm{\_gc}}$   (kpc), spatial coordinates $(x,y,z)$, velocity components $(v_x,v_y,v_z)$, and the galactocentric radial velocity,
\begin{equation}
v\_{\mathrm{radial\_gc}} = \frac{\vec{r}\cdot\vec{v}}{|\vec{r}|}\, .
\end{equation}
To characterize average motions, radial velocities were binned by $r{\mathrm{\_gc}}$ in 0.25~kpc intervals. For each bin, we computed the mean and its standard error ($\pm \sigma/\sqrt{N}$) and plotted $v\_ radial\_gc$ versus radius. 
\section{Results}
\textbf{Plots of radial trends.} Figure 1 show the mean galactocentric radial velocity ($v\_{\mathrm{radial} \mathrm{\_gc}}$) as a function of galactocentric distance ($r\_gc\_kpc$) in 0.25 kpc bins for representative sectors. Error bars indicate standard errors. In the first quadrant, the trend is strongly positive, reaching +20–25 km/s at larger radii. The second and third quadrants show weaker but consistently positive slopes, while the fourth quadrant displays a clear negative gradient (–15 to –28 km/s). These visual trends complement the statistical analysis and make the asymmetry between quadrants immediately apparent.
\begin{figure}[H]
    \centering
    \includegraphics[width=0.9\textwidth]{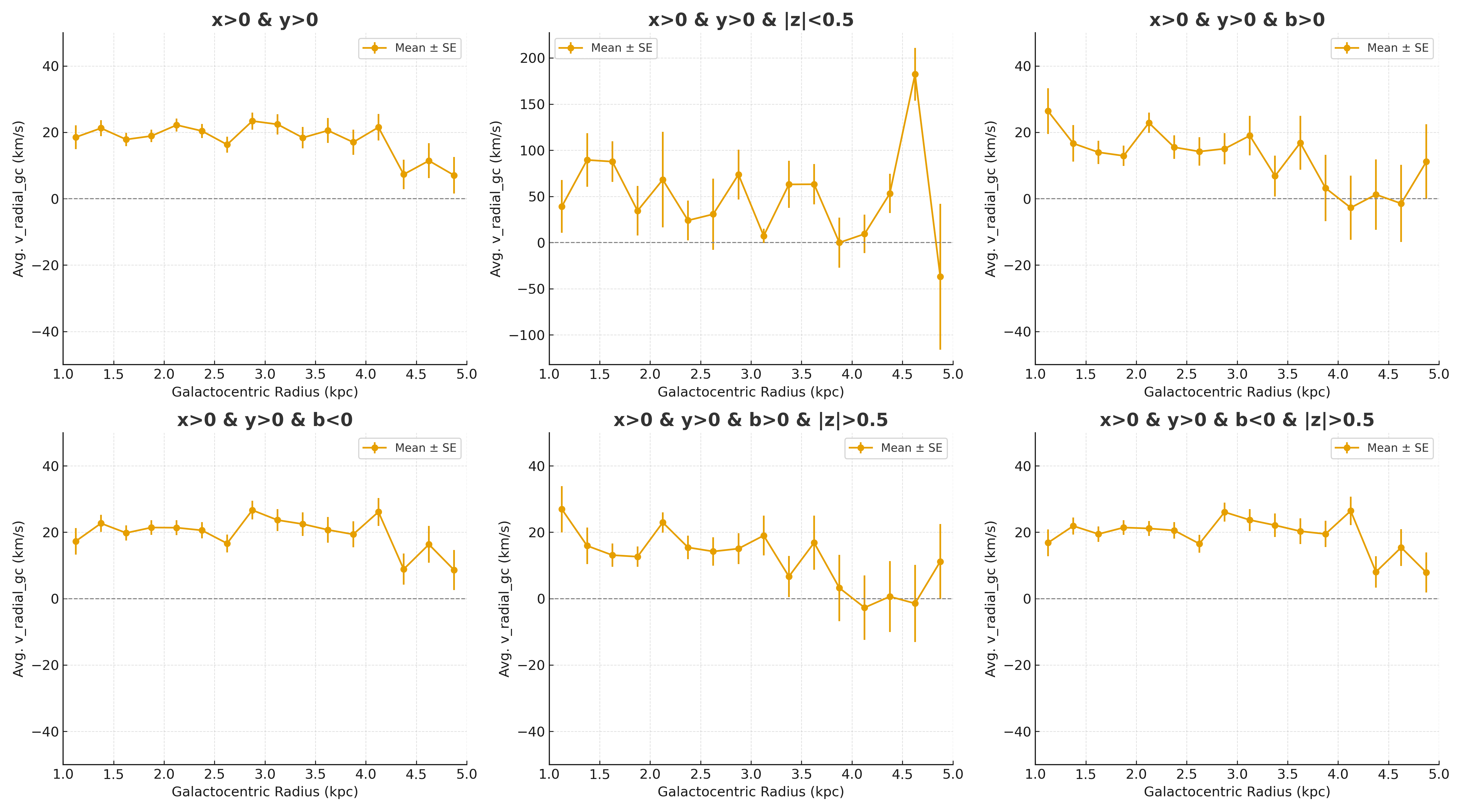}    \includegraphics[width=0.9\textwidth]{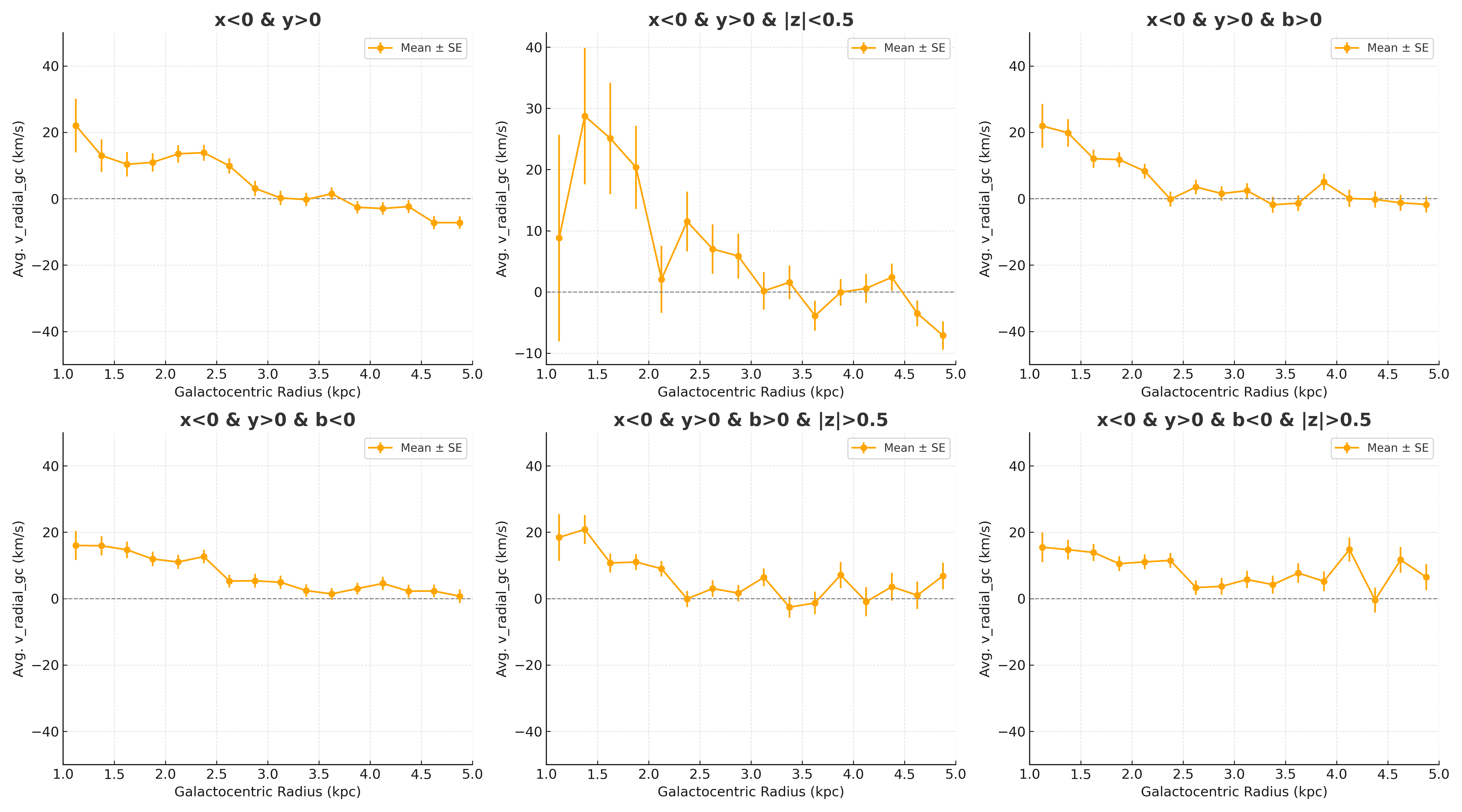}  
     \includegraphics[width=0.9\textwidth]{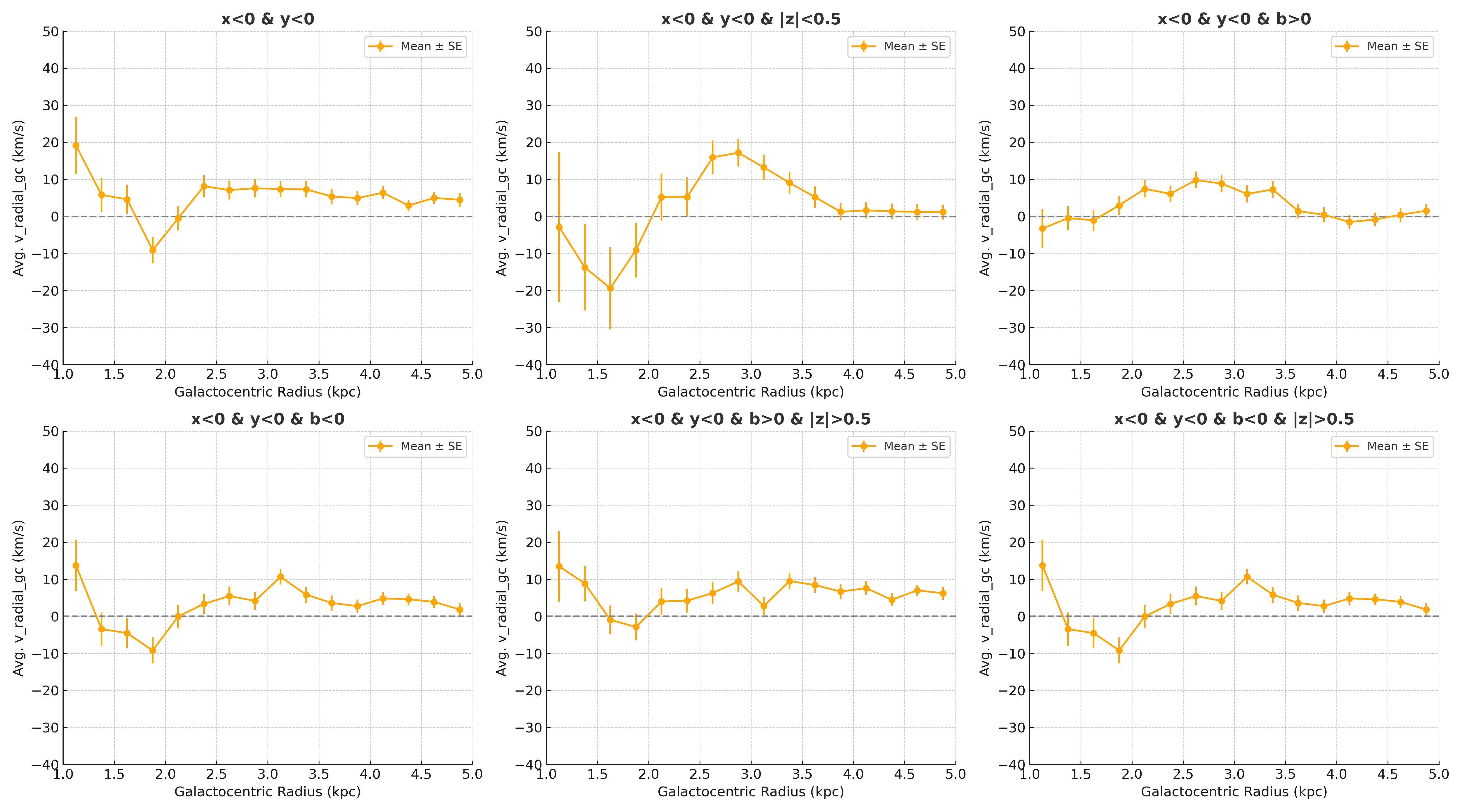}
\end{figure}
\begin{figure}[H]
    \centering
    \includegraphics[width=0.9\textwidth]{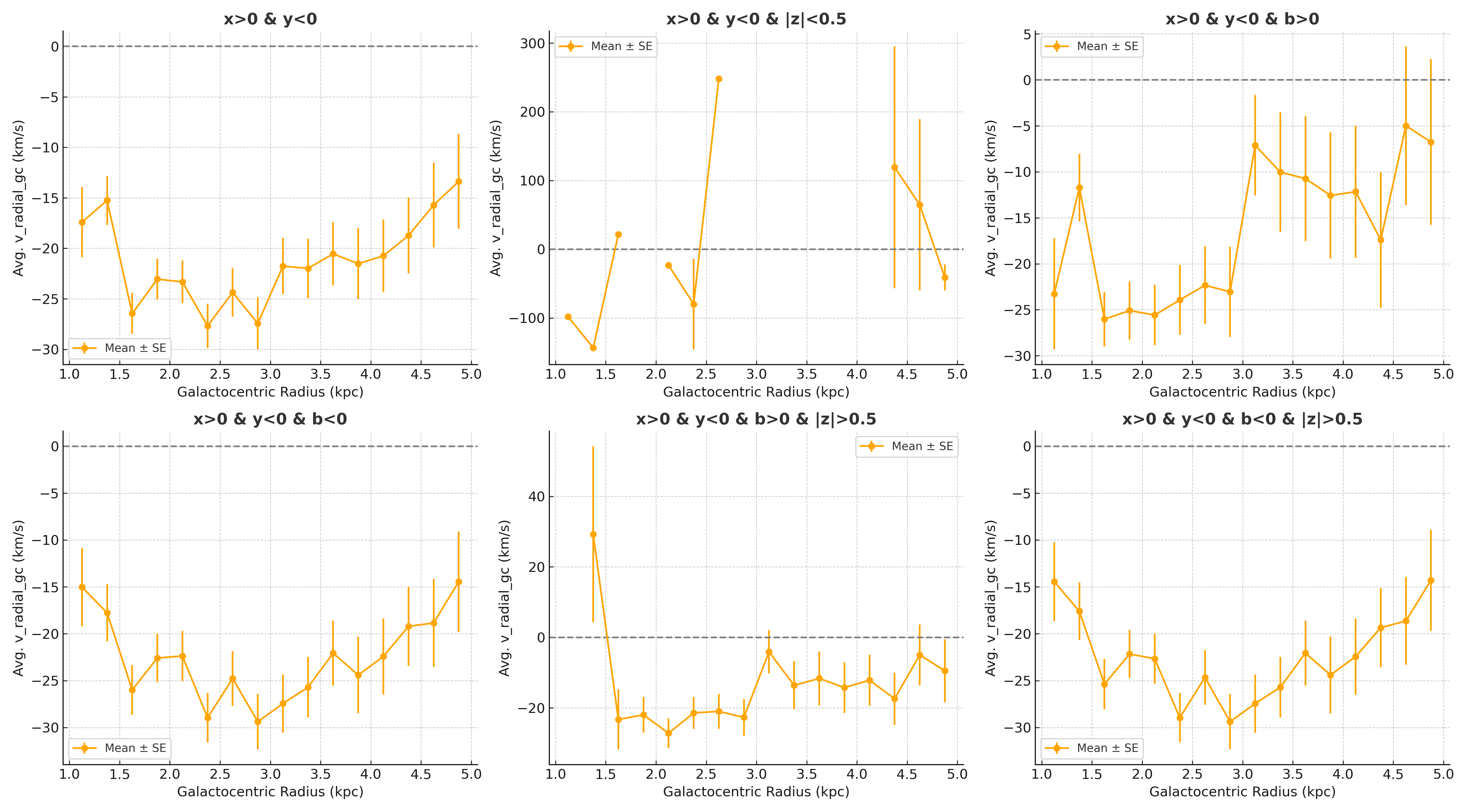}
    \includegraphics[width=0.9\textwidth]{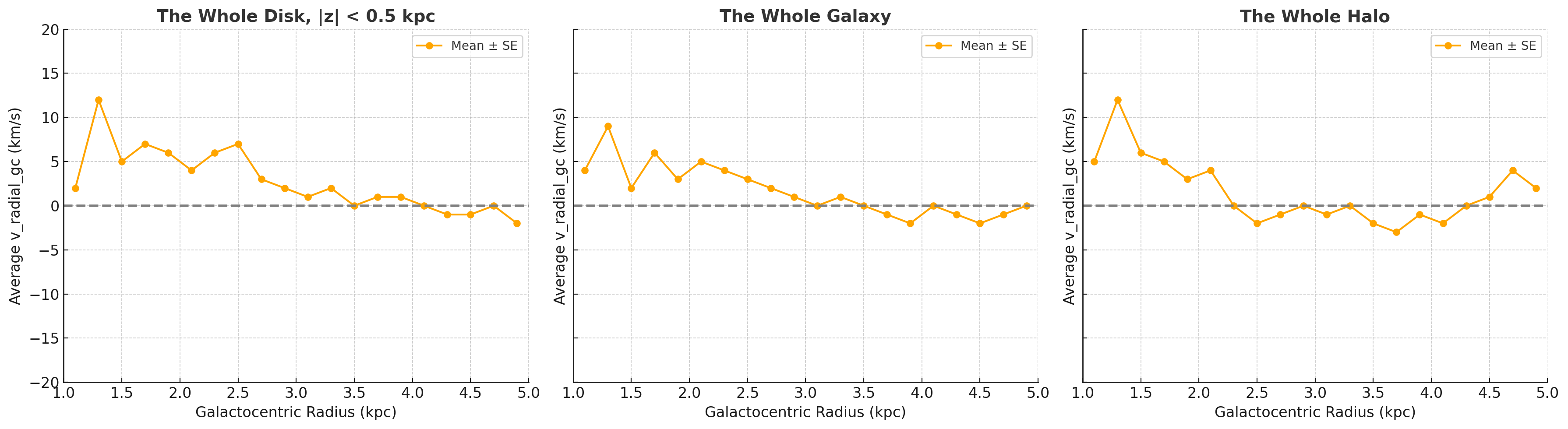}
    \caption{Plots of the mean galactocentric radial velocity ($v\_{\mathrm{radial} \mathrm{\_gc}}$) with error bars in 0.25 kpc bins, shown as a function of galactocentric distance ($r\_gc\_kpc$) for stars in different Galactic sectors. }
    \label{fig:quadrants}
\end{figure}

\textbf{Table~1 (Appendix 1)} summarizes the statistics of $v\_radial\_gc$ across 27 sectors within 5~kpc.

\textit{Quadrant analysis.} The first three quadrants ($x>0, y>0$; $x<0, y>0$; $x<0, y<0$) show consistently positive mean velocities.  

In the first quadrant, the mean velocity is 
$+17.2 \pm 0.7~\mathrm{km\,s^{-1}}$ ($N=27{,}141$, $p<0.001$), 
with higher values in the southern subsample 
($+19.2 \pm 0.8~\mathrm{km\,s^{-1}}$) and in the disk 
($|z|<0.5$ kpc), where the mean reaches 
$+49.9 \pm 7.3~\mathrm{km\,s^{-1}}$ (though based on only $N=204$ stars).  

The second quadrant yields smaller but significant positive values, 
$+2.5 \pm 0.6~\mathrm{km\,s^{-1}}$ ($N=22{,}818$, $p<0.001$).  
Subsamples range from $+1.6~\mathrm{km\,s^{-1}}$ in the disk 
($p\simeq 0.05$) to $+8.7~\mathrm{km\,s^{-1}}$ in the southern halo.  

The third quadrant also shows positive means, with the overall value 
$+5.1 \pm 0.6~\mathrm{km\,s^{-1}}$ ($N=20{,}617$, $p<0.001$).  
Subsamples fall between $+3.2$ and $+6.1~\mathrm{km\,s^{-1}}$.  

By contrast, the fourth quadrant ($x>0, y<0$) exhibits negative velocities.  
The mean is $-21.0 \pm 0.7~\mathrm{km\,s^{-1}}$ ($N=31{,}777$, $p<0.001$), 
with subsamples reaching $-18.5 \pm 1.2~\mathrm{km\,s^{-1}}$ (northern stars) 
and $-21.8 \pm 0.8~\mathrm{km\,s^{-1}}$ (southern stars).  
Halo components are also negative ($-17$ to $-22~\mathrm{km\,s^{-1}}$).  
The only exception is a very small disk subsample ($N=21$), 
where the mean is not statistically significant ($p=0.82$).

\textit{Disk, Galaxy, and halo averages.} When averaged over the Galactic disk ($|z|<0.5$ kpc), 
the mean galactocentric radial velocity is 
$+2.9 \pm 0.6~\mathrm{km\,s^{-1}}$ ($N=26{,}080$, $p<0.001$).  

For the whole Galaxy sample, the result is 
$+2.3 \pm 0.6~\mathrm{km\,s^{-1}}$ ($N=26{,}080$, $p<0.001$).  

The stellar halo also shows a smaller but significant expansion, 
with $+1.7 \pm 0.7~\mathrm{km\,s^{-1}}$ 
($N=18{,}911$, $p\simeq 0.02$).  

The ABV column in Table~1 reports the 
average binned velocity, computed in radial bins of 0.25 kpc, 
providing an independent confirmation of these results.

In the first quadrant ($x>0, y>0$), average binned velocities reach 
$12$-$49~\mathrm{km\,s^{-1}}$ (Table~1, Sectors~1--6), consistent with a 
clear outward motion.  
In the second and third quadrants ($x<0, y>0$ and $x<0, y<0$), 
average binned velocities remain positive but smaller, typically 
$\sim2$--$9~\mathrm{km\,s^{-1}}$ (Table~1, Sectors~7--18).  
By contrast, the fourth quadrant ($x>0, y<0$) shows consistently negative 
values, with binned means between $-13$ and $-15~\mathrm{km\,s^{-1}}$ 
(Table~1, Sectors~19--24).  

With the exception of one very small disk subsample 
(Sector~20, $N=21$, $p=0.82$), all sectors differ significantly from zero 
at the $p<0.05$ level.

\section{Discussion and Conclusion}
The sector-based analysis shows that 21 of 27 independent regions exhibit statistically significant outward velocities, while only five do not. Three quadrants yield positive mean values of +3 to +50 km s$^{-1}$, whereas the fourth quadrant displays negative velocities that transition mostly to positive beyond $\sim $ 6 kpc, (Figure 2) possibly reflecting the configuration of a localized activity zone and/or the influence of the Galactic bar (see e.g. \citep{Athanassoula1992};\citep{Fragkoudi2020}). This effect may arise from non-representative sampling of stars, for example due to the absorption of light from receding stars by the Galactic bar, or it may indicate that part of the activity zone does not coincide with the Galaxy’s dynamical center—a possibility that warrants further investigation. Despite this asymmetry, the global averages for the disk, halo, and full Galaxy remain significantly positive, indicating a net outward motion.

\begin{figure}
    \centering   \includegraphics[width=0.75\linewidth]{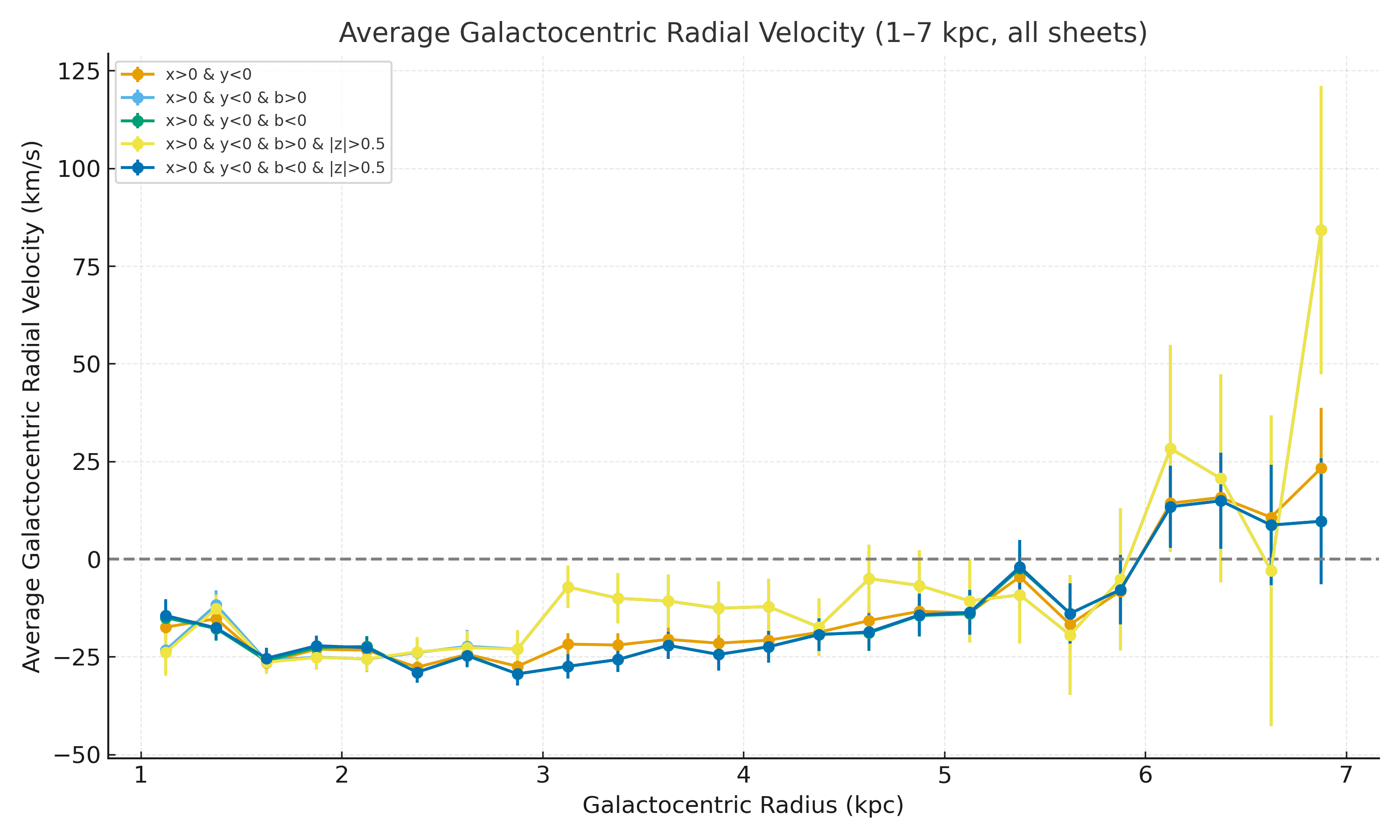}
    \caption{Fourth quadrant displays negative velocities that
transition to positive beyond
$\sim $ 6 kpc}
    \label{fig:placeholder}
\end{figure}

These results align with our earlier studies of globular clusters \citep{Karapetian2025a} and of stars along the south-north axis \citep{Karapetian2025b}, suggesting that both disk and halo populations may participate in large-scale expansion. If confirmed, such expansion would imply that the Galactic nucleus is dynamically active, injecting energy and matter into the surrounding system. This interpretation echoes the early hypothesis of V. Ambartsumian \citep{Ambartsumian1965}, later developed by Lynden-Bell \citep{LyndenBell1998}, that galaxies may originate from active central nuclei.

At the same time, several caveats must be noted.  Projection effects from Galactic rotation may contribute to errors and residual Gaia DR3 systematics cannot yet be ruled out. Therefore, the interpretation of expansion should be regarded as preliminary and requiring further confirmation. 

Nevertheless, if confirmed, these findings could offer a natural explanation for the rapid emergence of massive galaxies observed by JWST at ages of less than 1 Gyr, where accelerated growth driven by nuclear activity may be required. At the observed velocities, galaxies could reach sizes of about 10 - 20 kpc in less than 1 Gyr.

Future Gaia releases, together with complementary spectroscopic surveys, will be essential for testing and verifying this possibility.

\section*{Acknowledgments}
We thank the Gaia Collaboration and ESA for making DR3 data publicly available. We acknowledge use of \texttt{Astropy}, \texttt{astroquery}, and related Python tools.

\bibliographystyle{unsrt}
\bibliography{references}

\newpage
\appendix

\section*{Appendix 1: Table 1}

\begin{table}[!htbp]
\centering

\begin{tabular}{l l l l l l l l l}
 & Sector & N & Mean & StError & Median & Std & P\_ttest & ABV

 \\
1 & $x>0 \& y>0$ & 27141& 17.24& 0.68& 112.84& 19.69& 0.000& 17.78
\\
2 & $x>0 \& y>0 \& |z|<0.5$ & 204& 49.89& 7.34& 104.81& 54.08& 0.000& 49.25
\\
3 & $x>0 \& y>0 \& b>0$ & 8764& 11.83& 1.25& 116.66& 15.49& 0.000& 11.99
\\
4 & $x>0 \& y>0 \& b<0$ & 21088& 19.23& 0.76& 110.87& 21.23& 0.000& 19.54
\\
5 & $x>0 \& y>0 \& b>0 \& |z|>0.5$ & 8728& 11.65& 1.25& 116.70& 15.43& 0.000& 11.85
\\
6 & $x>0 \& y>0 \& b<0 \& |z|>0.5$ & 20892& 18.94& 0.77& 110.90& 20.95& 0.000& 19.20
\\
7 & $x<0 \& y>0$ & 22818& 2.50& 0.58& 88.23& 1.22& 0.000& 4.77
\\
8 & $x<0 \& y>0 \& |z|<0.5$ & 9499& 1.60& 0.82& 79.72& 1.85& 0.050& 6.09
\\
9 & $x<0 \& y>0 \& b>0$ & 21170& 3.76& 0.63& 91.20& 6.42& 0.000& 4.99
\\
10 & $x<0 \& y>0 \& b<0$ & 27831& 6.68& 0.53& 88.44& 5.70& 0.000& 7.16
\\
11 & $x<0 \& y>0 \& b>0 \& |z|>0.5$ & 14830& 5.38& 0.78& 95.01& 7.80& 0.000& 5.92
\\
12 & $x<0 \& y>0 \& b<0 \& |z|>0.5$ & 17780& 8.73& 0.70& 93.57& 8.71& 0.000& 8.73
\\
13 & $x<0 \& y<0$ & 20617& 5.14& 0.58& 83.14& 8.75& 0.000& 5.39
\\
14 & $x<0 \& y<0 \& |z|<0.5$ & 7715& 4.31& 0.81& 70.83& 7.01& 0.000& 1.77
\\
15 & $x<0 \& y<0 \& b>0$ & 23890& 3.17& 0.56& 86.87& 6.90& 0.000& 2.84
\\
16 & $x<0 \& y<0 \& b<0$ & 21246& 3.57& 0.57& 82.51& 6.98& 0.000& 2.99
\\
17 & $x<0 \& y<0 \& b>0 \& |z|>0.5$ & 19657& 6.09& 0.60& 84.24& 11.31& 0.000& 5.92
\\
18 & $x<0 \& y<0 \& b<0 \& |z|>0.5$ & 21246& 3.57& 0.57& 82.51& 6.98& 0.000& 2.99
\\
19 & $x>0 \& y<0$ & 31777& -20.97& 0.67& 119.42& -27.20& 0.000& -13.27
\\
20 & $x>0 \& y<0 \& |z|<0.5$ & 21& -7.99& 34.45& 157.88& -44.60& 0.819& 3.38
\\
21 & $x>0 \& y<0 \& b>0$ & 10648& -18.54& 1.17& 120.30& -23.14& 0.000& -7.58
\\
22 & $x>0 \& y<0 \& b<0$ & 22291& -21.83& 0.80& 119.19& -29.14& 0.000& -14.71
\\
23 & $x>0 \& y<0 \& b>0 \& |z|>0.5$ & 5511& -17.33& 1.64& 121.40& -23.19& 0.000& -14.88
\\
24 & $x>0 \& y<0 \& b<0 \& |z|>0.5$ & 22218& -21.72& 0.80& 119.07& -28.98& 0.000& -14.60
\\
25 & The Whole Disk, $|z| < 0.5$  & 26080& 2.87& 0.55& 88.90& 5.33& 0.000& 3.09
\\
26 & The Whole Galaxy $r\_{gc}\_{kpc} < 5$ & 26080& 2.25& 0.58& 93.24& 4.28& 0.000& 2.27
\\
27 & The Whole Halo $r\_{gc}\_{kpc} < 5$ & 18911& 1.71& 0.72& 98.98& 3.51& 0.018& 1.41
\\

\end{tabular}

\end{table}

\lstset{
  language=Python,
  basicstyle=\ttfamily\footnotesize,
  keywordstyle=\color{blue},
  commentstyle=\color{gray},
  stringstyle=\color{red},
  showstringspaces=false,
  breaklines=true,
  frame=single,
  numbers=left,
  numberstyle=\tiny\color{gray}
}

\section*{Appendix 2: Python Code}
The complete Python code (data extraction via ADQL, Gaia DR3 filtering, Galactocentric transformations, binning, and plotting).
\begin{lstlisting}[language=Python]
# -*- coding: utf-8 -*-
"""Main program.ipynb

Automatically generated by Colab.

"""

#!/usr/bin/env python3
# -*- coding: utf-8 -*-
"""
Gaia DR3 query (disk-wide example).
Note: run this script locally (not in Overleaf). Overleaf only typesets code.
Required once in your environment:
    pip install astroquery astropy pandas numpy openpyxl matplotlib
"""

from astroquery.gaia import Gaia
import pandas as pd
import numpy as np
from astropy.coordinates import SkyCoord, Galactocentric
import astropy.units as u
import matplotlib.pyplot as plt

TITLE_PREFIX = "(The Whole Disk, |z| < 0.5 kpc)"  # ASCII only (safe for LaTeX figures)

# ----------------------------- ADQL query -----------------------------
ADQL_QUERY = """
SELECT TOP 50000
  source_id, ra, dec, parallax, parallax_error, pmra, pmra_error, pmdec, pmdec_error,
  radial_velocity, radial_velocity_error, l, b, bp_rp, phot_g_mean_mag
FROM gaiadr3.gaia_source
WHERE
  parallax > 0
  AND l BETWEEN 0 AND 360
  AND b BETWEEN -90 AND 90
  AND parallax_over_error > 5
  AND ABS(radial_velocity) > 0
  AND ABS(radial_velocity / radial_velocity_error) > 5
  AND ABS(pmra / pmra_error) > 5
  AND ABS(pmdec / pmdec_error) > 5
"""

job = Gaia.launch_job_async(ADQL_QUERY)
df = job.get_results().to_pandas()
#print (df)

output_path = "fullsample.xlsx"  # define output_path before first use
df.to_excel(output_path, index=False)

#from google.colab import files
#files.download(output_path)

# Step 4: Convert to galactocentric coordinates and compute galactocentric velocity

def compute_galactocentric(row):
    try:
        coord = SkyCoord(
            ra=row['ra'] * u.deg,
            dec=row['dec'] * u.deg,
            distance=(1000.0 / row['parallax']) * u.pc,
            pm_ra_cosdec=row['pmra'] * u.mas/u.yr,
            pm_dec=row['pmdec'] * u.mas/u.yr,
            radial_velocity=row['radial_velocity'] * u.km/u.s
        )
        gc = coord.transform_to(Galactocentric())
        x = gc.x.to(u.kpc).value
        y = gc.y.to(u.kpc).value
        z = gc.z.to(u.kpc).value
        vx = gc.velocity.d_x.to(u.km/u.s).value
        vy = gc.velocity.d_y.to(u.km/u.s).value
        vz = gc.velocity.d_z.to(u.km/u.s).value
        r_vec = gc.cartesian.xyz.to(u.kpc).value
        v_vec = gc.velocity.d_xyz.to(u.km/u.s).value
        v_radial_gc = np.dot(r_vec, v_vec) / np.linalg.norm(r_vec)
        r = np.linalg.norm(r_vec)
        return pd.Series([r, x, y, z, vx, vy, vz, v_radial_gc])
    except:
        return pd.Series([np.nan]*8)

# Apply transformation
cols = ['r_gc_kpc', 'x_kpc', 'y_kpc', 'z_kpc', 'vx_kms', 'vy_kms', 'vz_kms', 'v_radial_gc']
df[cols] = df.apply(compute_galactocentric, axis=1)

# Step 5: Filter for stars
df_disk = df[
  (df['r_gc_kpc'] > 1)&
  (df['r_gc_kpc'] < 5)&
  (df['z_kpc'].abs() < 0.5)
  #(df['x_kpc'] < 0)&
  #(df['y_kpc'] < 0)
  ]


#print(df[['x_kpc', 'y_kpc', 'z_kpc']].describe())

# Step 6: Save to Excel with full kinematics
output_path = "disk_stars_gaia_dr3.xlsx" # redefine output_path before second use
df_disk.to_excel(output_path, index=False)

print(df_disk.columns)

# Step 7: Plot z-distribution
plt.figure(figsize=(8, 5))
plt.hist(df_disk['z_kpc'], bins=50, color='skyblue', edgecolor='k')
plt.xlabel("Z (kpc)")
plt.ylabel("Number of Stars")
plt.title(f"z-distribution {TITLE_PREFIX}")
plt.grid(True)
plt.legend()
plt.tight_layout()
plt.show()

# Step 7b: Plot galactocentric velocity vs. galactocentric distance
plt.figure(figsize=(8, 5))
plt.scatter(df_disk['r_gc_kpc'], df_disk['v_radial_gc'], s=5, alpha=0.5)
plt.axhline(0, color='gray', linestyle='--')
plt.xlabel("Galactocentric Radius (kpc)")
plt.ylabel("Galactocentric Radial Velocity (km/s)")
plt.title(f"Galactocentric Radial Velocities {TITLE_PREFIX}")
plt.grid(True)
plt.tight_layout()
plt.show()

# Step 7c: Plot average galactocentric velocity in radial bins with error bars
bin_edges = np.arange(0, 5, 0.25)
bin_centers = 0.5 * (bin_edges[1:] + bin_edges[:-1])
grouped = df_disk.groupby(pd.cut(df_disk['r_gc_kpc'], bin_edges))
binned_means = grouped['v_radial_gc'].mean()
binned_stds = grouped['v_radial_gc'].std()
binned_counts = grouped['v_radial_gc'].count()
binned_errors = binned_stds / np.sqrt(binned_counts)

PLOT_COLOR = "orange"
plt.figure(figsize=(8, 5))
plt.errorbar(bin_centers, binned_means, yerr=binned_errors,
             fmt='o-', capsize=4, label='Mean ± Error', color=PLOT_COLOR)
# Optional: keep an extra line if you like, but use same color for consistency
# plt.plot(bin_centers, binned_means, linestyle='-', color=PLOT_COLOR, alpha=0.9)
plt.axhline(0, color='gray', linestyle='--')
plt.xlabel("Galactocentric Radius (kpc)")
plt.ylabel("Average Galactocentric Radial Velocity (km/s)")
plt.title(f"Average galactocentric velocity in radial bins 0.25 kpc {TITLE_PREFIX}")
plt.grid(True)
plt.legend()
plt.tight_layout()
plt.show()

 # Step 8: Save both full and disk samples in a single Excel file with two sheets
combined_excel_path = "gaia_full_and_disk_stars.xlsx"



# Step 9: Download the combined Excel file
from google.colab import files
files.download(combined_excel_path)

\end{lstlisting}

\end{document}